\documentstyle[11pt]{article}
\begin{document}
\author{A.Ya. Kazakov \\
Laboratory of Quantum Information, \\
State University of AeroSpace Instrumentation, \\
190000, Bol'shaya Morskaya, 67, S.-Petersburg, Russia\\
e-mail: a\_kazak@mail.ru}
\title{Dense coding and safety of quantum communications}
\maketitle

\section{Introduction}
Safety of quantum communications is, in the first place, the
result of the quantum nature of the signal, sending by the sender
Alice to the receiver Bob \cite{Bennett}, \cite{Zeilinger},
\cite{Steane}. But when we discuss the safety of the quantum
communications as a whole we have to take into account all objects
taking place in the  process of creation and communication of
information.  One of such object is the sending station, which is
classical one. As such it can be eavesdropped by classical means
and this fact can reduce sharply the safety of quantum channel. In
this note we demonstrate, that dense coding and quantum properties
of the channel give the possibility to raise the safety of
classical sending station against the eavesdropping.

 \section{Main considerations}
We discuss in what follows the next situation. We consider a
quantum channel, which contains a quantum system $L=L_A\bigotimes
L_B$, $dimL_A=n, dimL_B=p$, where quantum system $L_A$ is
initially in the hands of Alice. There is an {\it initial} quantum
state $\Psi_{0}\in L$, which is distributed between Alice and Bob.
Alice influences on her part of wave function and Bob gets {\it
outcome } state $(U\bigotimes I_B)\Psi_{0}$, where $U$ is an
unitary matrix describing the interaction of Alice with quantum
system $L_A$. This unitary matrix corresponds to the state of the
classical sending station and plotter can use classical means in
order to get some information about $U$.

  At first we consider the following question arising in this situation:
  how large is the set of the outcome states generated by Alice,
$\hat{L}=\{(U\bigotimes  I_B)\Psi_{0}\}$, where $U$ - all possible
unitary matrices acting in $L_A$. The unitary matrix does not
change the norm of wave function, so $dim_{\textbf{R}}\hat{L}\leq
 dim_{\textbf{R}}L-1=2np-1$. When the equality is realized it is said
 that there is the {\it dense coding}. Evidently, that equality can not be
 realized for any initial state $\Psi_{0}$. If, for instance,
  $\Psi_{0}=|\phi >|\varphi  >$, $|\phi >\in L_A,
|\varphi >\in L_B$, then $\hat{L}=L_A\bigotimes |\varphi >$. So,
we describe at first such states  $\Psi_{0}$, that condition
\begin{equation}\label{q1}
dim_{\textbf{R}}\hat{L}= dim_{\textbf{R}}L-1=2np-1.
\end{equation}
is valid.

Further, let us discuss the relation between the dimensions of our
varieties. Unitary matrix $U$ acts in the linear space $L_A$,
whose complex dimension is $n$, so $dim_R\{U\}=n^2$. It is
possible situation, when
\begin{equation}\label{q2}
  K=n^2-2np>0.
\end{equation}
 In this case the unitary matrices $E(\kappa )$, $\kappa\in
{\textbf{R}}^{K}$ exist, which do not change the outcome state,
\begin{equation}\label{q3}
  (E(\kappa)\bigotimes I_B)\Psi_{0}=\Psi_{0}.
\end{equation}
for all $\kappa $.  Such unitary matrices we call {\it conserving
matrices }. If unitary matrix $U_{1}$ corresponds to some sending
information, we conclude, that the same communication can be
created by unitary matrix $U_{1}E(\kappa )$ for any $\kappa $.
Choosing this parameter arbitrarily,  for instance, by chance, we
can reduce the value of information about state of classical
sending station, in other words, we raise the safety of channel of
communication as a whole.

   In what follows we discuss the structure of the initial state
   $\Psi_{0}$, and describe on this base the set of conserving matrices.
It will be shown below, that condition (\ref{q2}) can be softened.
 \section{The structure of the initial state}
Consider now initial states $\Psi_{0}$, when condition (\ref{q1})
holds. Let set $|\varphi _{k}>, k=1,2,...,p$ is a basis in $L_B$.
Then
\begin{equation}\label{q5}
  \Psi_{0}=\sum _{k}|\phi _{k}>|\varphi _{k}>,
\end{equation}
and $|\phi _{k}>\in L_A$ for all $k$. We prove here, that vectors
$|\phi _{k}>$ are linear independent. Really, let for some $k_{0}$
É and constants $\lambda _{k}$ $|\phi _{k_{0}}>=\sum_{k\neq
k_{0}}\lambda _{k}|\phi _{k}>$. Then
\begin{equation}\label{q7}
\Psi_{0}=\sum _{k\neq k_{0}}|\phi _{k}>[|\varphi _{k}>+\lambda
_{k}|\varphi _{k_{0}}>],
\end{equation}
The set of vectors $|\varphi _{k}>+\lambda _{k}|\varphi _{k_{0}}>,
k\neq k_{0}$, is not a basis in the space $L_B$, because there are
 $p-1$ such vectors in number. We conclude, that set $\{(U\bigotimes
 I_B)\Psi_{0}\}$ does not coincide with $L/R$, so condition
 (\ref{q1}) is not valid.

It follows from this consideration, that initial state $\Psi_{0}$
contains in its expansion $|\varphi _{k}>$ at any $k=1,2,...,p$.

So, dense coding realizes only in the case, when initial state for
 $\Psi_{0}$ is {\it entangled} one.
\section{Set of the conserving matrices}
Here we get the explicit description of conserving matrices, which
satisfy the condition  (\ref{q3}). It follows from this condition,
that
\begin{equation}
\label{s1}
E(\kappa )|\phi _{k}>=|\phi _{k}>, k=1,2,...,p.
\end{equation}
Let $H \subset L_A$ is a linear span of vectors $|\phi _{k}>,
k=1,2,...,p$, $H^{\perp}$ is its orthogonal addition. Then
$E(\kappa )\mid_{H}=I\mid_{H}, E(\kappa )H^{\perp}=H^{\perp}$. So,
 the set of unitary operators satisfying condition (\ref{q3}) coincides
 with the set of unitary operators acting on $H^{\perp}$. The dimension of
 the last set of operators equals $(n-p)^2$. Here we suppose, that
condition (\ref{q2}) is changed by the more soft variant: $n>p$.

The set of conserving unitary matrices can be described in the
following way. Assuming, that matrix $E(\kappa )$ can be
differentiating on $\kappa $ and using the described above
structure of initial state $\Psi_{0}$, we obtain:
\begin{equation}\label{q9}
  D|\phi _{k}>=0, k=1,2,...,p,
\end{equation}
where skew-Hermitian matrix $D=[dE/d\kappa](\kappa =0)$ acts in
space $L_A$. This $n\times n$-matrix can be described as
\begin{equation}
D=\left(
\begin{array}{cccc}
i\delta _1 & \alpha _{12} & \alpha _{13} & ... \\
-\overline{\alpha _{12}} & i\delta _2 & \alpha _{23} & ... \\
-\overline{\alpha _{13}} & -\overline{\alpha _{23}} & i\delta _3 & ... \\
... & ... & ... & ...
\end{array}
\right)
\label{matrix}
\end{equation}

Transferring to the real parameters $\mid \phi _k^{(r)}>,\mid \phi
_k^{(i)}>,\alpha _{st}^{(r)},\alpha _{st}^{(i)}$, so that $\mid
\phi _k>=\mid \phi _k^{(r)}>+i\mid \phi _k^{(i)}>,\alpha
_{st}=\alpha _{st}^{(r)}+i\alpha _{st}^{(i)}$, we obtain the
following relations for $\alpha_{kn}^{r},\alpha_{kn}^{i}, \delta
_{k}$:

\begin{equation}
\left(
\begin{array}{cccc}
0 & \alpha _{12}^{(r)} & \alpha _{13}^{(r)} & ... \\
-\alpha _{12}^{(r)} & 0 & \alpha _{23}^{(r)} & ... \\
-\alpha _{13}^{(r)} & -\alpha _{23}^{(r)} & 0 & ... \\
... & ... & ... & ...
\end{array}
\right) \mid \varphi _k^{(r)}>-\left(
\begin{array}{cccc}
\delta _1 & \alpha _{12}^{(i)} & \alpha _{13}^{(i)} & ... \\
\alpha _{12}^{(i)} & \delta _2 & \alpha _{23}^{(i)} & ... \\
\alpha _{13}^{(i)} & \alpha _{23}^{(i)} & \delta _3 & ... \\
... & ... & ... & ...
\end{array}
\right) \mid \varphi _k^{(i)}>=0,  \label{z1}
\end{equation}

\begin{equation}
\left(
\begin{array}{cccc}
\delta _1 & \alpha _{12}^{(i)} & \alpha _{13}^{(i)} & ... \\
\alpha _{12}^{(i)} & \delta _2 & \alpha _{23}^{(i)} & ... \\
\alpha _{13}^{(i)} & \alpha _{23}^{(i)} & \delta _3 & ... \\
... & ... & ... & ...
\end{array}
\right) \mid \varphi _k^{(r)}>+\left(
\begin{array}{cccc}
0 & \alpha _{12}^{(r)} & \alpha _{13}^{(r)} & ... \\
-\alpha _{12}^{(r)} & 0 & \alpha _{23}^{(r)} & ... \\
-\alpha _{13}^{(r)} & -\alpha _{23}^{(r)} & 0 & ... \\
... & ... & ... & ...
\end{array}
\right) \mid \varphi _k^{(i)}>=0,  \label{z2}
\end{equation}
As was shown before, this system of $2np$ real linear algebraic
equations for $n^2$ real unknown parameters
$\alpha_{kl}^{r},\alpha_{kl}^{i}, \delta _{k}$ has $S=(n-p)^2$
fundamental solutions. In accordance with (\ref{matrix}) we can
construct skew-Hermitian matrix for any fundamental solution,
$D_s, s=1,2,...,S$. Then unitary matrix $E_s(\gamma _s)=\exp
\left[ \gamma _sD_s\right] $, where $\gamma _s$ is an arbitrary
real number, is the conserving matrix. So, matrix
$U=\prod_{s=1}^SE_s(\gamma _s)$ depending on $S$ real parameters
$\gamma _s , s=1,2,...,S$ is the general conserving unitary
matrix, where ordering on $s$ is arbitrary but fixed. Note, that
for the construction of the conserving unitary matrix $U$ it is
necessary to find the set of fundamental solutions for system
(\ref{z1},\ref{z2}), using the linear independent vectors $\mid
\phi _k>, k=1,2,...,p$ (which are components of expansion of the
initial entangled state $\Psi_{0}$).

  \section{Conclusion}
We have shown, that dense coding under condition (\ref{q2})
between the dimensions of the quantum systems $L_A$ and $L_B$
leads to the existence of the unitary matrix conserving the
outcome states. This fact results in the ambiguity between the
states of the classical sending station and outcome state - there
are different states of the sending station for the same outcome
state. The presence of such "additional" degrees of freedom for
the sending station gives the possibility to "hide" the sending
information and to raise the safety of the quantum channel as a
whole. Let us note the next features.

 1. The described possibility to increase  the safety of channel
 is connected with its quantum properties.

 2. The construction of the conserving matrices depends on the information about
 the initial entangled state  $\Psi_{0}$. In a certain sense this quantum state is
 a key, distributed between Alice and Bob, for the set of conserving matrices.

At the realization of quantum systems $L_A$ and $L_B$ with help of
set of qubits the number of additional degrees of freedom for the
conserving matrices can be enough large for the moderate numbers
of qubits. Really, let $L_A$ contains $m$ qubits whose dimension
is $d$, $L_B$ contains $q$ the same qubits, so $n=d^m, p=d^q$,
$S=(d^m-d^q)^2$. If  $d=2, m=3, q=1$, then $S=36$.

I thank V. Gorbachev, N. Shekhunova  and A. Trubilko for
stimulating discussions.

\end{document}